\documentclass[doublecol]{epl2} 

\title{Femtosecond laser pulse shaping for enhanced ionization}
\shorttitle{Femtosecond laser pulse shaping for enhanced ionization}

\author{A. Castro\inst{1}\thanks{E-mail: \email{alberto@physik.fu-berlin.de}}
        \and E. R{\"a}s{\"a}nen\inst{1,2} \thanks{E-mail: \email{erasanen@jyu.fi}}
        \and A. Rubio\inst{3} 
        \and E. K. U. Gross\inst{1}}
\shortauthor{A. Castro \etal}

\institute{                    
  \inst{1} Institut f{\"u}r Theoretische Physik 
           and European Theoretical Spectroscopy Facility (ETSF),
           Freie Universit{\"a}t Berlin, Arnimallee 14, D-14195 Berlin, Germany\\
  \inst{2} Nanoscience Center, Department of Physics, 
           University of Jyv{\"a}skyl{\"a}, FI-40014 Jyv{\"a}skyl{\"a}, Finland\\
  \inst{3} Nano-Bio Spectroscopy group and European Theoretical Spectroscopy Facility
           (ETSF), Dpto.  de Fisica de Materiales, Universidad del Pais Vasco UPV/EHU, 
           Centro Mixto CSIC-UPV/EHU and DIPC, Av. Tolosa 72, E-20018 San Sebasti{\'{a}}n, Spain\\
}


\abstract{
  We demonstrate how the shape of femtosecond laser pulses can be
  tailored in order to obtain maximal ionization of atoms or
  molecules. For that purpose, we have overlayed a direct-optimization
  scheme on top of a fully unconstrained computation of the
  three-dimensional time-dependent Schr{\"{o}}dinger equation.  The
  procedure looks for pulses that maintain the same total length and
  integrated intensity or fluence as a given pulse that serves as an
  initial guess. It allows, however, for changes in frequencies
  -- within a certain, predefined range -- and overall
  shape, leading to enhanced ionization.  We illustrate the scheme by
  calculating ionization yields for the H$_2^+$ molecule when
  irradiated with short ($\approx 5$ fs), high-intensity laser
  pulses.
}

\begin{document}

\maketitle

\section{Introduction}

When atoms or molecules are irradiated with laser fields that are
intense enough to induce nonlinear effects, a wealth of fascinating
phenomena may be observed~\cite{posthumus-2004}.  This applies even to
deceitfully ``uninteresting'' systems such as the simplest molecule,
H$_2^+$~\cite{giusti-suzor-1995}: above-threshold or tunneling
ionization~\cite{gibson-1997}, bond softening~\cite{bucksbaum-1990},
bond hardening (light induced bound states or vibrational
trapping)~\cite{fransinski-1999}, charge resonance enhanced
ionization~\cite{zuo-1993, chelkowski-1996}, above threshold
dissociation~\cite{giusti-suzor-1990}, high harmonic
generation~\cite{plummer-1995}, etc.

This very same complexity in the molecular reaction, however, is what
permits to envision the possibility of \emph{controlling} molecules
with short (femtosecond time scale) and intense (10$^{11} - 10^{15}$
W/cm$^2$) laser pulses~\cite{shapiro-2003}. The short durations allow
for \emph{coherent control}: the systems evolve uncoupled to the
environment, and can be steered towards the desired outcomes without
relying in the more traditional control parameters, i.e., average,
thermodynamic functions such as the temperature. The high intensities
trigger the strongly nonlinear, even non-perturbative, response of the
systems. An essential ingredient to realize the molecular control is
the capability of shaping the laser pulses -- a technological area
that has witnessed spectacular advances in the recent
years~\cite{weiner-2000}.

Yet this complexity implies the need for challenging theoretical
models. Not surprisingly, ionization is the first and most studied
process, mainly because it could already be studied for
atoms~\cite{protopapas-1997}, and because in this intensity regime it
almost always occurs, be it accompanied or not by other
phenomena. Even in the absence of influence from nuclear dynamics, the
ionization of molecules is significantly more complex than that of
atoms, due to the electron emission from different atomic
centers~\cite{muth-bohm-2000, litvinyuk-2005}. Two rather successful
models for molecular ionization that have recently been suggested are
the so-called molecular-orbital strong-field
approximation~\cite{muth-bohm-2000} and the molecular extension of
the Ammosov-Delone-Kraine (ADK)
approximation~\cite{tong-2002}. However, these approaches are
insufficient as \emph{general} tools~\cite{awasthi-2008}.

A common feature of the approaches mentioned above is the use of the
single-active-electron (SAE) approximation. One-electron molecular
systems such as the hydrogen molecular ion H$_2^+$ are therefore
perfect candidates to isolate the error introduced by the SAE
approximation from further simplifications. Electron correlation
originates difficult and interesting phenomena such as non-sequential
ionization~\cite{walker-1994}.

In order to properly investigate the interaction of short and intense
laser fields with molecules, one needs to perform explicitly
time-dependent calculations, even if it might imply a heavy
computational burden. Calculations of this kind, that propagate the
time-dependent Schr{\"{o}}dinger equation (TDSE), have been presented
for H$_2^+$ in the past, for example with the purpose of understanding
the presence of maxima in the ionization yield for particular
internuclear separations~\cite{enhanced-ionization}, or in order to
disentangle the relationship between ionization and
dissociation~\cite{ionization-dissociation,
  chelkowski-1996}. Recently, Selst{\o} {\em et
  al.}~\cite{selsto-2005} and Kjeldsen {\em et
  al.}~\cite{kjeldsen-2006} have reported calculations on the
orientation dependence of the ionization yield -- lifting the commonly
used assumption of a molecular axis parallel to the light
polarization.

In this work we take a further step, and focus on the possibility of
theoretically designing, via fixed-nuclei three-dimensional (3D) TDSE
calculations, laser pulses able to control (in particular,
significantly enhance) the ionization yields, taking H$_2^+$ as an
example system.  Some recent experimental breakthroughs on this area
have triggered our interest. For example, Suzuki {\em et
  al.}~\cite{suzuki-2004} demonstrated the control of the multiphoton
ionization channels of I$_2$ molecules by making use of a pulse
shaping system capable of varying in time the polarization
directions. Simultaneously, Brixner {\em et al.}~\cite{brixner-2004}
have made use of a similar polarization-shaping system to enhance
ionization yields of diatomic molecules (K$_2$).  Our focus is,
however, on linearly polarized ultrashort pulses ($\approx$ 5~fs), so
rapid that the nuclear movement does not play a role during the pulse
action -- in contrast to the studies in which the ionization
is studied as the internuclear distance changes, leading to possible
resonances.

\section{Methodology}

The optimization problem could be formulated in the language of
quantum optimal-control theory (QOCT)~\cite{qoct}. It consists of a
set of equations -- along with various suitable, iterative algorithms
that solve them -- whose solution provides an optimized \emph{control
  field} that typically maximizes a \emph{target operator} $\hat{O}$.
In order to enhance ionization, one would just define ${\hat{O}}$ as
the projection onto unbound states, or, alternatively, the identity
minus the projection onto the bound states:
\begin{equation}
\label{eq:operator}
\hat{O} = \hat{1} - \sum^{\rm bound}_i \vert\varphi_i\rangle\langle\varphi_i\vert\,.
\end{equation}
However, we have experienced numerical difficulties when attempting to
solve the QOCT equations for this particular operator: The
forward-backward propagations that must be performed in order to solve
the QOCT equations proved to be, for our particular implementation,
numerically unfeasible when using the operator given in
Eq.~(\ref{eq:operator}) to define the target. This was due to the
appearance of fields with unrealistically high frequencies and/or
amplitudes. We believe that the reason lies in the fact that the
backward propagation must be performed after acting with the operator
$\hat{O}$ on the previously propagated wave function. This eliminates
the smooth, \emph{numerically friendly} part of the wave function,
enhancing, on the contrary, the high frequency components. This
procedure is repeated at each iteration, eventually making the
propagation impossible. We do not claim, however, that any other
numerical implementation will not be able to successfully cope with
this problem.

Therefore, we have employed and present here, a {\emph{direct}}
optimization scheme, which is in fact much closer in spirit to the
techniques utilized by the experimentalists~\cite{rabitz-2000}.  In
this scheme, we construct a merit function by considering the
expectation value of the operator defined in Eq.~(\ref{eq:operator})
at the end of the propagation:
\begin{equation}
F(x) =
\langle\Psi_x(T)\vert\hat{O}\vert\Psi_x(T)\rangle\,,
\end{equation}
where $x$ is the set of parameters that define the laser pulse, and
$\vert\Psi_x(T)\rangle$ is the wave function that results from
performing the propagation with the laser determined by $x$, at the
final time $T$.  Of course, the sum over the bound states has to be
truncated; for the calculations presented below, we find it sufficient
to include the lowest ten states. The merit function is calculated by
performing consecutive TDSE propagations: The resulting function
values are fed into a recently developed derivative-free algorithm
called NEWUOA~\cite{newuoa}. This algorithm seeks the maximum of any
merit function $F(x)$ depending on $N$ variables $x$, and does not
necessitate the gradient $\nabla F$. It is very effective for $N$
larger than ten and smaller than a few hundred, which is the case
considered here. In all our runs, we necessitated around two hundred
iterations to converge the ionization yields to within 1\%.

The system and the TDSE propagations are modeled in our homegrown {\tt
  octopus} code~\cite{octopus}. We represent the wave functions on a
real-space rectangular regular grid, and fix the nuclear position at
their equilibrium distance. The small length of the pulses used here
justifies this simplification. We perform calculations setting the
polarization direction both parallel to the molecular axis and
perpendicular to it. The size of the simulation box is selected large
enough to ensure that very little of the electronic density has
reached the grid boundaries at the end of the laser pulse.
Nevertheless, we add absorbing boundaries to remove this charge; if
the propagation is pursued after the pulse, part of the density will
``abandon'' the simulation box; the remaining integrated density
should approach (as it does) one minus the ionization probability
calculated as the expectation value of Eq.~(\ref{eq:operator}).

The laser pulse is taken in the dipolar approximation, and represented
in the length gauge. The temporal shape of the pulse is given by a
function $f(t)$, which we expand in a Fourier series:
\begin{equation}
f(t) = f_0 + \sum_{n=1}^N 
\left[ f_n \sqrt{\frac{2}{T}}\cos(\omega_n t)
+ g_n \sqrt{\frac{2}{T}}\sin(\omega_n t) \right]\,,
\end{equation}
with $\omega_n =  2\pi n/T$.
In order to ensure a physically meaningful laser pulse~\cite{madsen}, 
we must have
$\int_0^T {\rm d}t f(t)=0$, which implies $f_0=0$.
Moreover, we must have $f(0)=f(T)=0$, where $T$ is the total
propagation time. This poses the following constraint:
\begin{equation}
\label{eq:condition-zero}
\sum_{n=1}^N f_n = 0\,.
\end{equation}
The sum over frequencies is truncated according to physical
considerations: Any pulse shaper must have a predefined range of
frequencies it can work with. The feasibility of the numerical scheme
depends on the possibility of truncating the previous expression at a
reasonably low number. In the cases considered here, due to the short
duration of the pulses, we obtain no more than around 20 degrees of
freedom by setting the maximum frequency to one Hartree.

Evidently, by increasing the intensity of a pulse one can enhance the
ionization yield. Our wish is to improve this yield by changing the
pulse shape, and not simply by lasing with larger intensity.
Therefore, to ensure the {\emph{fairness}} in the optimization search,
we constrain the search to laser pulses whose time-integrated
intensity (fluence) is predefined to some value $F_0$:
\begin{equation}
F_0 = \int_0^T\!\!\!\! f^2(t)\,{\rm d}t = \sum_{n=1}^N (f_n^2+g_n^2)\,.
\end{equation}
The search space $\lbrace f_n,g_n\rbrace$ is thus constrained to the
hyper-sphere defined by the previous equation. But we must add the
condition given by Eq.~(\ref{eq:condition-zero}), which further
restricts the search space to a hyper-ellipsoid. By performing the
appropriate unitary transformation, this can be brought again into a
hypersphere:
\begin{equation}
F_0 = \sum_{n=1}^{2N-1} \xi_n^2\,.
\end{equation}
This equal-fluence condition can be guaranteed if we perform a new
transformation to hyperspherical coordinates:
\begin{eqnarray}
\nonumber
\xi_1 & = & F_0^{1/2} \cos(\theta_1)\,,
\\\nonumber
\xi_2 & = & F_0^{1/2} \sin(\theta_1) \cos(\theta_2)\,,
\\\nonumber
\dots & = & \dots
\\\nonumber
\xi_{2N-2} & = & F_0^{1/2} \sin(\theta_1) \dots \sin(\theta_{2N-3}) \cos(\theta_{2N-2})\,,
\\
\xi_{2N-1}   & = & F_0^{1/2} \sin(\theta_1) \dots \sin(\theta_{2N-3}) \sin(\theta_{2N-2})\,.
\end{eqnarray}
The set of angles $\lbrace \theta_j\rbrace_{j=1}^{2N-2}$ constitute the
$2N-2$ variables that define the search space for the optimization
algorithm.

\section{Results}

The initial laser field before the optimization is a linearly
polarized eight-cycle pulse having a sinusoidal envelope, fixed peak
intensity, and wavelength of $\lambda=400$ nm -- a typical value for
frequency-doubled Titanium-sapphire lasers.  Correspondingly, the
initial frequency is $\omega_0= 0.114$ Ha, and the pulse
length is $5.3$ fs.  The maximum allowed
frequency of the {\em optimized} pulse is set to $\omega_{\rm
  max}=2\,\omega_0$.  The pulse polarization is fixed to be parallel
or perpendicular to the molecular axis. During the QOCT procedure, the
polarization and the fluence $F_0$ are kept fixed, but the {\em peak}
intensity may change from the initial value, which is selected in the
range $I=0.5, 0.75, \ldots, 2\times 10^{15}$ W/cm$^2$.

\begin{figure}
\centerline{\includegraphics[width=0.7\columnwidth]{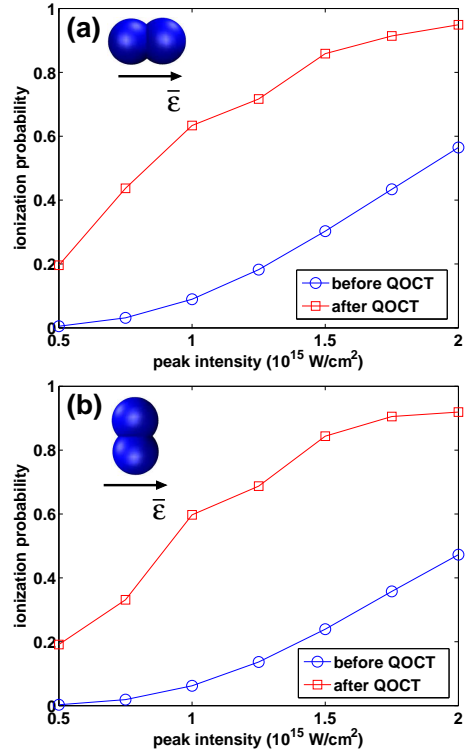}}
\caption{(color online). 
Ionization probability for the initial pulse (circles) and for 
the optimized pulse (squares) as a function of the peak
intensity of the initial pulse. The polarization
of the pulse is (a) parallel and (b) perpendicular to
the molecular axis.
}
\label{fig1}
\end{figure}

Figure~\ref{fig1} 
shows the ionization probabilities as a function of
the peak intensity (of the initial guess pulse) for the initial and
optimized pulses polarized parallel (a) and perpendicular (b) to the
molecular axis, respectively. Overall, the pulse optimization leads to
a significant increase in the ionization. As expected, the ionization
yield is slightly larger for pulses polarized parallel to the
molecular axis. 

\begin{figure}
\centerline{\includegraphics[width=0.99\columnwidth]{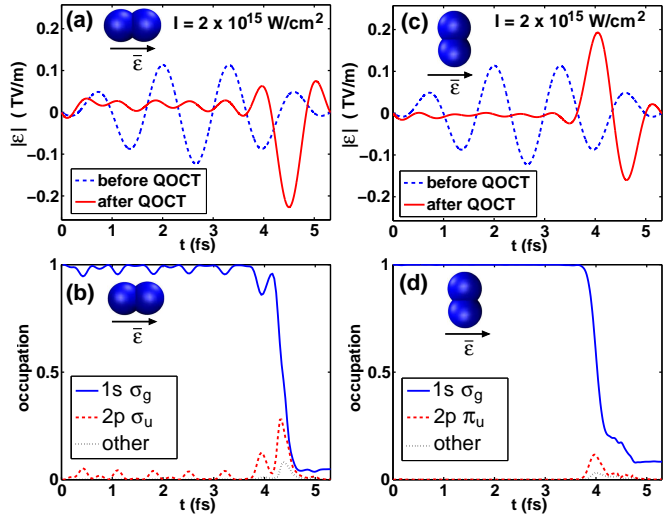}}
\caption{(color online).
(a) Initial and optimized pulses (parallel polarization) and (b) the occupation of 
selected single-electron states in the optimized ionization process,
when $I=2\times 10^{15}$ W/cm$^2$. (c-d) Same as (a-b) but for
perpendicular polarization. 
}
\label{fig2}
\end{figure}

\begin{figure}
\centerline{\includegraphics[width=0.99\columnwidth]{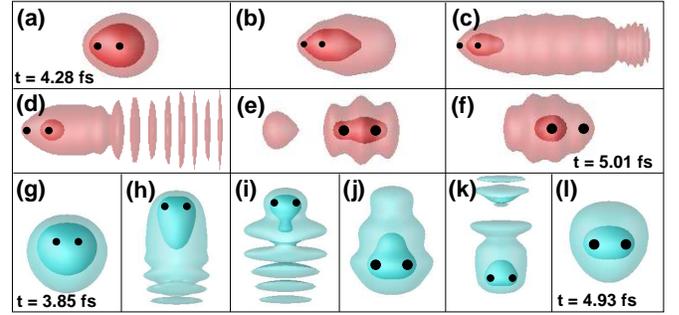}}
\caption{(color online).  (a-f) Snapshots of the electron densities in
  the optimized ionization process when the pulse polarization is
  parallel to the molecular axis and the intensity is $I=2\times
  10^{15}$ W/cm$^2$. The black dots mark the positions of the
  nuclei. The time-steps between snaphots are all equal. (g-l) The
  same for a pulse polarized perpendicular to the molecular axis.  }
\label{fig3}
\end{figure}

To get more insight into the optimized ionization process, we plot in
Fig.~\ref{fig2} 
the initial and optimal laser pulses and the occupations of some
single-electron states during the pulse interaction. The peak
intensity of the initial pulse is $I=2\times 10^{15}$ W/cm$^2$.
Optimized pulses of both parallel (a) and perpendicular (c)
polarization have large peaks near the end of the pulse. According to
the corresponding occupations shown in Figs.~\ref{fig2}(b) and (d),
these amplitude peaks account for almost all of the ionization: During
the peaks the ground-state occupations rapidly collapse.  The 2p
$\sigma_{\rm u}$ (2p $\pi_{\rm u}$) excited state contributes to the
process to a small extent in the parallel (perpendicular) case,
whereas the other states are involved by a nearly negligible fraction;
overall, no excited bound states contribute significantly. Hence,
within the constraints set here for the laser pulse, the optimal
ionization of H$_2^+$ is a direct process obtained by focusing most of
the available pulse energy in a very short time frame -- though
keeping the integrated total field at zero in accordance with
Maxwell's equations (see Ref.~\cite{madsen}). The electron densities
during the ionization process are visualized in Fig.~\ref{fig3}
for both parallel (a) and perpendicular (b) polarizations.

The previous optimizations have produced rather ``uninteresting''
solutions. It is a well known fact that in a short intense laser
pulse, most of the ionization occurs during the peaks in the electric
field.  Therefore, the optimizations have just attempted to create
short, intense bursts of light. This fact can be understood if we
consider the process happening in the quasi-static, tunneling regime,
in which the total ionization can be approximated by considering at
each moment in time the static ionization rate that corresponds to the
electric amplitude. This ionization rate is nonlinear, and it is much
larger at the electric field peaks, which therefore cause most of the
ionization. Note, however, that the cases discussed above lie in an
intermediate regime between the tunneling and the multi-photon regime
-- the Keldysh parameter, $\gamma$, is of the order of one [The
  Keldysh parameter $\gamma$ is defined as $\sqrt{\vert
    E_I\vert/2U_p}$, where $E_I$ is the ionization potential of the
  system, and $U_p$ is the pondemorotive energy, given in atomics units
  by $(E_0/2\omega)^2$, $E_0$ being the peak intensity of the electric
  field, and $\omega$ the pulse frequency.  Since our optimized lasers
  do not have a single frequency -- not even necessarily a dominant
  one, we can only speak of approximate Keldysh parameters.]

Alternatively, one can explain the simplicity of the pulses
considering that the maximum allowed frequency, $2\,\omega_0 =
0.228$~Ha, is smaller than any resonance transition energy from the
ground state. As a consequence, the system does not significantly
populate these states, and the only ionizing channel is direct
transition to the continuum.

\begin{figure}
\centerline{\includegraphics[width=0.9\columnwidth]{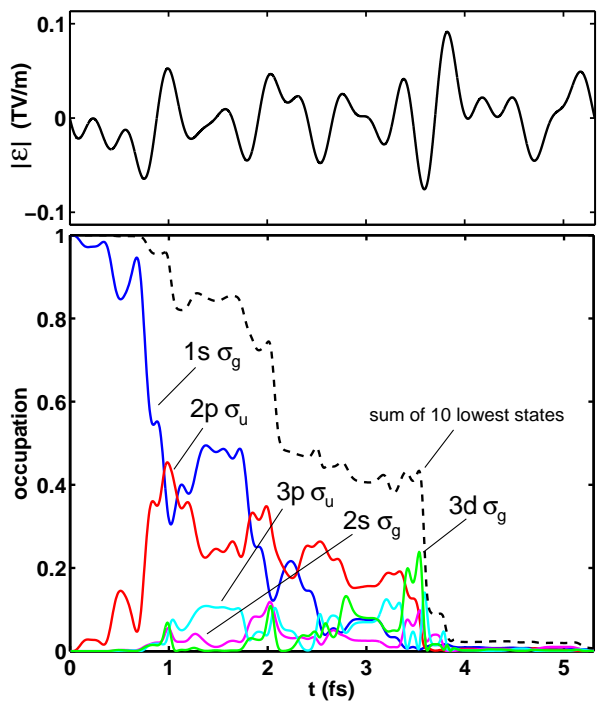}}
\caption{
(color online). Upper panel: Optimized laser pulse for the ionization
when the cutoff frequency is $4\,\omega_0$ (see text) and the intensity
is fixed to $0.5 \times 10^{15}$ W/cm$^2$. Lower panel: Occupation of
a few lowest states during the pulse interaction.
}
\label{fig4}
\end{figure}

The picture changes significantly, however, if we allow for a larger
cutoff frequency. First, this increases the value of the Keldysh
parameter associated with the process, which may change the regime
from a more quasistatic to a more of a multi-photon-like character.
Secondly, the excited bound states are now accessible for
single-photon transitions. For example, Fig.~\ref{fig4} displays
results obtained for $4\,\omega_0$. The intensity is here set to
$0.5\times 10^{15}$~W/cm$^2$. Doubling the cutoff frequency of the
search space has a significant effect in the total ionization yield:
Now we obtain 0.99 for the ionization probability, whereas in the
first optimization the yield was 0.20 (see Fig.~\ref{fig1}, top panel,
first point in the series). Note that the initial yield before any
optimization was only 0.005.

Moreover, the manner in which the ionization occurs with a larger
cutoff frequency is qualitatively very different.  Figure~\ref{fig4}
displays the evolution of the occupation of some of the bound
states. The first excited state ($2{\rm p}\,\sigma_{\rm u}$) plays a
significant role, which can be understood because the transition
energy from the ground state is now accessible in the field search
space. 
In addition, a couple of other lowest states contribute in the
ionization process in an ascending order as a function of time. It
should be noted, however, that only the $\sigma$ orbitals, where the
nodes are perpendicular to the polarization axis, participate in the
transitions, and $\pi$ orbitals, for example, are not allowed due to a
different symmetry. As a consequence of the involvement of several
states in the ionization process, the structure of the optimized laser
pulse shown in the upper panel of Fig.~\ref{fig4} is much more
complicated than the single-burst fields obtained in the previous
calculations.

\section{Conclusions}
In conclusion, we have shown with three-dimensional time-propagations
of the time-dependent Schr\"odinger equation how the precise temporal
shape of a short intense laser pulse may affect significantly the
total ionization yield of H$_2^+$ at fixed internuclear
separations. Moreover, we have employed a gradient-free optimization
technique to find the laser pulse that enhances ionization. This
optimization can be constrained in different ways, accounting for the
limitations of physical sources -- not all frequencies and intensities
are available, and not all possible shapes can be constructed with the
state-of-the-art pulse shapers (although the technology improves at a
phenomenal rate). The results will differ depending on these
constraints: The optimized laser pulse may be the single burst of
electric field that one would expect by considering a process in the
tunnelling regime, or a field with a more complicated structure that
drives the system through intermediate states -- the ionization can be
enhanced by resonant transitions.  Whereas in the former case it would
be easy to design intuitively pulses that maximize the ionization, in
the latter an optimization algorithm such as the one presented in
this work is necessary.

\acknowledgments We thank M. F{\o}rre and L. Madsen for helpful
discussions. We acknowledge funding by the European Community through
the e-I3 ETSF project (INFRA-2007-1.2.2: Grant Agreement Number
211956).  AR acknowledges support by the Spanish MEC
(FIS2007-65702-C02-01), "Grupos Consolidados UPV/EHU del Gobierno
Vasco" (IT-319-07), CSIC, the Barcelona Supercomputing Center, "Red
Espanola de Supercomputacion" and SGIker ARINA (UPV/EHU). ER
acknowledges support by the Academy of Finland. AC and EKUG
acknowledge support from the Deutsche Forschungsgemeinschaft within
the SFB 658, and EKUG acknowledges hospitality from KITP-Sta. Barbara.

\end{document}